%Paper: 9204008
%From: Michael Flatte <flatte@sbphy.physics.ucsb.edu>
%Date: Tue, 21 Apr 92 11:03:45 -0700

%%%%%%%%%%%%%%%%%%%%%%%%%%%%%%%%%%%%%%%%%%%%%%%%
%                                              %
% the tex file with text and figure captions   %
% begins here.  two postscript files, for      %
% figures 1 and 2, follow.  they are           %
% separated by headings like this one.         %
% these should be separated into their own     %
% files and then printed separately.           %
%                                              %
%%%%%%%%%%%%%%%%%%%%%%%%%%%%%%%%%%%%%%%%%%%%%%%%

\magnification=\magstep1

\hsize=6.5truein
\vsize=8.9 truein
\baselineskip=\normalbaselineskip \multiply\baselineskip by
2
\def\beginlinemode{\endmode\begingroup\parskip=0pt
\obeylines\def\\{\par}\def\endmode{\par\endgroup}}

\def\beginparmode{\endmode\begingroup
\def\endmode{\par\endgroup}}

\def\endpage{\vfill\eject}

\def\raggedcenter{\leftskip=4em plus 12 em
\rightskip=\leftskip \parindent=0pt \parfillskip=0pt
\spaceskip=.3333em \xspaceskip=.5em \pretolerance =9999
\tolerance=9999 \hyphenpenalty=9999 \exhyphenpenalty=9999 }

\let\endmode=\par{\obeylines\gdef\
{}}

\overfullrule=0pt

\medskipamount=7.2pt plus2.4pt minus2.4pt

\parskip=\medskipamount

\def\refto#1{$^{#1}$}

\gdef\refis#1{\indent\hbox to 0pt{\hss#1.~}}

\gdef\journal#1, #2, #3, 1#4#5#6{{\sl #1~}{\bf #2}, #3, (1#4#5#6)}

\def\prb{\journal Phys. Rev. B, }

\def\etal{{\it et al.}}

\def\prl{\journal Phys. Rev. Lett., }

\def\pr{\journal Phys. Rev., }

\def\rmp{\journal Rev. Mod. Phys., }

\rightline{UCSBTH-92-13}
\rightline{cond-mat/9204008}

\null\vskip 3pt plus 0.2fill \beginlinemode
\raggedcenter {\bf Image of the Energy Gap Anisotropy in the Vibrational
Spectrum of a High Temperature Superconductor}

\vskip 3pt plus 0.2fill Michael E. Flatt\'e

\vskip 3pt plus 0.2fill {\sl Department of Physics\\
University of California\\Santa Barbara, CA  93106--9530}

\vskip 3pt plus 0.3fill \beginparmode \narrower We present a new
method of determining the anisotropy
of the gap function in layered high-$\rm T_c$ superconductors.
Careful inelastic neutron scattering measurements at low
temperature
of the phonon dispersion curves in the
(100) direction in $\rm La_{1.85}Sr_{.15}CuO_4$
would determine whether the gap is predominately
s-wave or d-wave.
We also propose an experiment to
determine the gap at each point on a quasi-two-dimensional
Fermi surface.


\footline={\hss\tenrm\folio\hss}

\endpage\beginparmode

The anisotropy of the energy gap at the Fermi surface
contains clues to the pairing mechanism
operating in a superconductor.  An s-wave gap\refto{1}
successfully describes phonon-mediated pairing, which causes
superconductivity in ordinary metals.  A
p-wave gap\refto{2} describes spin-flip-mediated pairing in
helium-3.
Current work suggests that the pairing in heavy fermion
materials is mediated by anti-ferromagnetic
fluctuations and is predominately d-wave\refto{3}.
Theoretical work on high-$\rm T_c$ materials has produced
theories too diverse and numerous to list, even as classes,
but all must address the form of the gap function.
Therefore we expect that accurate measurements of the gap
would provide
a strong test which any theory of high-$\rm T_c$ materials must
pass.

The experiments
available to look for gap anisotropy provide limited
information.  These experiments can be divided into two
classes: those which provide some average of the gap magnitude
over
all or a large part of the (normal) Fermi surface, and those which identify
nodes in the gap on the Fermi surface.

Reflectivity\refto{4}, tunneling current\refto{5},
ultrasonic attenuation\refto{6} and thermal
conductivity\refto{7} measurements, as well as a
variety of other transport measurements, yield some average value of the
gap magnitude.  In addition to their fundamental limitation
-- providing an average over the gap -- there are other difficulties
characteristic of each technique, which are discussed in the
above references.  Identifying nodes on the Fermi surface is
done by measuring the temperature dependence of quantities
such as ultrasonic attenuation\refto{8},
specific heat\refto{9}, magnetic penetration depth\refto{10}
or magnetic relaxation rate\refto{11}.
Unfortunately, the exponential temperature dependence, that
would rule out nodes, and the power-law dependence, that
would indicate nodes, are difficult to distinguish.  And even if
the evidence for a node is convincing, its location is hard
to determine because of the previously-mentioned averaging
effects.

We present a new method of analyzing measurements of phonon
dispersion curves
which should determine whether the gap in
a high-$\rm T_c$ material is well described by a
predominately s-wave or d-wave
function. This method should also locate any nodes on the
Fermi surface.
In addition, recognizing that gaps which are approximately
s, p, or d-wave are
special cases of an anisotropic gap, we propose an
experiment which can determine the gap magnitude everywhere on the
Fermi surface.
Both of these techniques require accurate measurements of
phonon
dispersion curves, which can now be made with inelastic
neutron scattering\refto{12}.

The gap at the Fermi surface
and the Fermi surface shape completely
determine
the minimum energy
required for a phonon
to decay into two Bogoliubov quasiparticles. For phonon
wavelengths smaller than the coherence length, the minimum-energy
excitation is the creation of both quasiparticles
on the Fermi surface.  The positions on the Fermi surface
where the quasiparticles are created can be found by placing
the phonon momentum vector in the Brillouin zone so that its
head and tail are on the Fermi surface (see Figure 1).
The quasiparticles are created at these points.
The Fermi
surface shape is described by a function $\vec
k_F(\theta)$, which is the vector whose tail is at the
center of the Brillouin zone, whose head lies on the Fermi surface,
and whose angle to a fixed axis is $\theta$.
Given the phonon momentum and the function $\vec k_F(\theta)$,
one can find the angles $\phi$ and
$\tilde\phi$ where $\vec q$'s head and tail lie.
Therefore
we have the following equation, which should be viewed as
determining $\phi$ and $\tilde\phi$, given $\vec q$ and the
function $\vec k_F(\theta)$:
$$\vec q = \vec k_F(\phi) - \vec k_F(\tilde\phi).\eqno(1)$$
The placement of the phonon momentum, and thus $\phi$ and
$\tilde\phi$,
are usually two-fold degenerate for a two-dimensional
Fermi surface such as $\rm La_{2-x}Sr_xCuO_4$'s. If the
Fermi surface and the gap magnitude are inversion symmetric,
this degeneracy is trivial and $\phi$ can be considered
defined only from $0$ to $\pi$. $\rm La_{2-x}Sr_xCuO_4$'s
Fermi surface is inversion-symmetric, and an
inversion-symmetric gap magnitude conforms to current
expectations\refto{13}.  We will discuss situations with
non-trivial degeneracies at the end of this Letter.

The minimum, or
threshold, energy
is the sum of the gap magnitude at the two points $\vec k_F(\phi)$ and
$\vec k_F(\tilde\phi)$
on the Fermi surface.
These two points are uniquely determined by $\vec q$ and the
function $\vec k_F(\theta)$, so the
threshold energy, denoted $\tilde \Delta$, can be considered as a
function of $\vec q$ and a functional of $\vec k_F$.  In the
remainder of this Letter any dependence on
$\vec k_F$ will be considered implicit.  This
paragraph, therefore, can be summarized by the following
equation:
$$\tilde\Delta(\vec q) = |\Delta(\phi)| +
|\Delta(\tilde\phi)|\eqno(2)$$
where $\Delta(\phi)$ is the gap at the point on the Fermi
surface where the head of $\vec k_F(\phi)$ is.
As will be discussed later, $\tilde\Delta$ depends
on $\vec q$ in markedly different ways if the gap is d-like
instead of s-like.

We will now present a way to determine the shape of
the surface defined by
$\tilde\Delta(\vec q)$
in energy-momentum
space.
Phonons with energies below the surface,
which we will call
the threshold surface,
should in general live longer than those above it.
However, a variety of other influences on the phonon
lifetimes blur this distinction.  Fortunately, near and just
above the
threshold surface, the effect on the phonon lifetime due to the
newly opened quasiparticle channel becomes strongly amplified.
The lifetimes and frequencies of phonons are usually
plotted as a function of momentum magnitude, $q$, with the
direction $\hat q$ fixed.  With $\hat q$ fixed,
$\tilde\Delta(q)$ defines a threshold line instead of a
surface.  Our proposal to determine $\tilde\Delta(\vec q)$
is to observe anomalous behavior in
the lifetimes and frequencies
of the phonon branches which cross the threshold line.

The transition from below to above the threshold line
is sharp.  Consider for illustrative purposes a weak-coupling
BCS model where other
influences on the phonon lifetime are approximated by a constant
lifetime $\tau_o$.  More realistic models will be discussed
at the end of this Letter.
We will follow a phonon dispersion curve which crosses the
threshold line, beginning with a $q_1$ such that
$\hbar\omega(q_1)<\tilde\Delta(q_1)$ and concluding with
a $q_3$ such that
$\hbar\omega(q_3)>\tilde\Delta(q_3)$.  At the
threshold momentum $q_2$, the phonon frequency is on the
threshold line:
$\hbar\omega(q_2)=\tilde\Delta(q_2)$.
This model predicts
the lifetime is $\tau_o$ for $q_1<q<q_2$.  The
lifetime at $q=q_2$
will vanish. At and slightly above the threshold line, the
lifetime will be proportional
to $(q-q_2)^{1/2}$ and eventually will approach $\tau_o$ if $q_3$
is sufficiently larger than
$q_2$.

This anomalous behavior suggests that points on the threshold
surface can be identified by measuring the lifetimes of
phonons and determining where in $\vec q,\omega$ space they
drop precipitously.  These points satisfy the equation
$$\hbar\omega(\vec q) = |\Delta (\phi)| +
|\Delta(\tilde\phi)| = \tilde\Delta (\vec q).\eqno(3)$$
Equations (1) and (3) will be referred to as the kinematic
equations.
Each time a
phonon dispersion curve crosses the threshold surface, one
can determine a point on the surface.  Since there are a
finite number of dispersion curves, $\tilde\Delta(\vec q)$ will be
measureable  at a few points.
This does not determine the function $\Delta (\theta)$ completely, but
suffices to distinguish predominately
s-wave, p-wave, or d-wave superconductivity.

Figure 2 is a plot of the threshold line $\tilde\Delta(q)$
in the (100) direction for a modified tight-binding Fermi
surface\refto{14}
(shown in Figure 1) with electron occupation $0.85$.
 This is an
approximation to the Fermi surface of
$\rm La_{1.85}Sr_{.15}CuO_4$, which has the highest $\rm T_c$ for
$\rm La_{2-x}Sr_xCuO_4$.  Plotted is the threshold line for a
d-wave gap
$$\Delta(\theta)=\Delta_o\left[{\rm cos}\left(k_F(\theta){\rm cos}\theta\right)
-{\rm cos}\left(k_F(\theta){\rm sin}\theta\right)\right]\eqno(4)$$
and an s-wave gap
$\Delta(\theta) = \Delta_o$. We draw the reader's attention
to the node in $\tilde\Delta(q)$ near the half-way point to
the zone boundary.  The only way to have a node in the
threshold line is to have that phonon momentum connect two
nodes on the Fermi surface (shown as $\vec q_n$ on Figure
1).  So, this node in the threshold surface is a signature
of nodes in the gap function.

The number of (100) phonon branches which cross the threshold in
the d-wave case is approximately equal to the number of zone
center phonons with energies less than twice the maximum gap.
High-$\rm T_c$ superconductors, therefore, are good
materials to examine for this signature because of their
large gap.

The gap function magnitude can be determined {\it everywhere} on
the Fermi surface by performing a
different experiment.  For a particular phonon branch, the
kinematic equations have
solutions which form curves in $\vec q, \omega$ space.
Assume that, given all these curves, at least one solution
to the kinematic equations exists for each value of $\phi$.
It requires quite a pathological case for this not to be
true.
For each $\phi$ pick one solution and identify it with $\phi$.
Knowing a solution to the kinematic equations means one knows
the values of the quantities $\omega(\vec q)$ and
$\tilde\phi$.  So these quantities can now be considered
functions of $\phi$.
Define the Fourier coefficients of a function of $\phi$ as
follows:
$f(\phi) = \sum_n f_{1n} {\rm cos}(n\phi)+f_{2n} {\rm sin}(n\phi)$.
Equation (3), the energy equation,
becomes the following set of coupled linear
equations for the Fourier coefficients $|\Delta|_{in}$ of
the gap magnitude:
$$\hbar\omega_{in} = |\Delta|_{in} +
\sum_{jm}c_{injm}|\Delta|_{jm}\eqno(5)$$
where
$$ c_{1n2m}={1\over \pi}\int_0^{\pi}d\phi
{\rm sin}(m\phi){\rm cos}(n\tilde\phi(\phi))\eqno(6)$$
and the other $c_{injm}$ involve the other three
combinations of trigonometric functions.
The inversion symmetry of the gap magnitude forces
$|\Delta|_{in}=0$ for $n$ odd.   Matrix inversion of
Equation (5) determines the $|\Delta|_{in}$, and thus the
gap magnitude everywhere on the Fermi surface.

For a two-dimensional Fermi surface that is not inversion
symmetric, or that is not convex, there are sometimes two or
more non-trivial solutions to Equation (1) for
$\vec q$.
The kinematic equations no
longer always
uniquely determine $\phi$ and $\tilde\phi$ in terms of $\vec
q$, therefore,
the threshold energy $\tilde\Delta(\vec
q)$ becomes the
minimum of $|\Delta(\phi)|+|\Delta(\tilde\phi)|$ for all
possible $\phi,\tilde\phi$ pairs.
This newly defined threshold surface will still have
structure if the gap is anisotropic.  In particular, nodes
in the gap function will be readily visible, with the
threshold surface plunging to zero for a momentum connecting
two nodes. The inversion procedure which culminates in
Equation (5) remains unchanged if one is assured that any
$\vec q,\omega$ solution to the kinematic equations which is
used in the procedure only has a trivial degeneracy.

For a three-dimensional Fermi surface this inversion
procedure is no longer possible. In this case for a
momentum $\vec q$,
Equation (2)
has an infinite number of solutions for $\phi$ and $\tilde\phi$,
which are now solid angles.
A direct algorithm for evaluating $|\Delta(\theta)|$
using $\vec k_F(\theta)$ and the solutions to the
kinematic equations is no longer possible because a fit of
$|\Delta|_{in}$ now requires multiparameter root finding.
However the concept of a threshold surface remains valid and
the nodes of it still indicate a momentum which joins gap nodes
on the Fermi surface.

Only a simple BCS model has been presented, but
a calculation including
strong electron-phonon coupling and Coulomb
interactions\refto{15,16} indicates that
the lifetime retains large anomalous features.
In this calculation, the lifetime at $q=q_2$
does not reach zero and the discontinuity at
this threshold momentum is smoothed.
The rise of the phonon lifetime for $q>q_2$, however, is much faster
than in the weak-coupling case.  Because of these competing
effects it is difficult to determine whether
the transition from below to above the
threshold surface would be
easier or harder to observe in this more realistic
model than in the weak-coupling model.

The coherence factor in the electromagnetic
response function provides a minor complication.
For a system with the d-wave gap of Equation (4), this
factor vanishes for a phonon
whose wavevector is parallel to the (110) direction and
whose head and tail lie on the Fermi surface.  For
such a phonon, therefore, the
minimum-energy excitation does not contribute to its
self-energy.
This phonon's lifetime would still decrease upon crossing
the threshold line, but would not be discontinuous at the
threshold.
We do not expect this
behavior could be observed, considering current experimental
capabilities.  The solution to this problem is
simply to consider threshold lines in other directions, as
done in Figure 2. Only particular
phonon directions are affected, therefore
this complication does not affect
the inversion procedure described above for the gap magnitude.

Since an extremely sensitive probe of surface
phonons exists in thermal energy inelastic helium
scattering\refto{17}, we remark that similar arguments to
those presented in this paper may lead to gap measurement
techniques involving surface phonons.

\vskip 0.25truein
{\raggedcenter Acknowledgments \par}\nobreak\vskip
0.25truein\nobreak
We would like to acknowledge extremely useful conversations
with W. Kohn and D.J. Scalapino.
This work is supported by the National Science Foundation through Grant
No. DMR87-03434 and
the U.S. Office of Naval Research through Grant
No. N00014-89-J-1530.

\filbreak\vskip 0.5truein{\raggedcenter REFERENCES
\par}\nobreak\vskip 0.25truein\nobreak
\refis{1} J. Bardeen, L.N. Cooper, J.R. Schrieffer, \pr 108, 1175, 1957.

\refis{2} A.J. Leggett, \rmp 47, 331, 1975.
P. W. Anderson, W.F. Brinkman, in {\it The
Physics of Liquid and Solid Helium}, K. H. Bennemann and J.
B. Ketterson, Eds. (Wiley, New York, 1978), part 2, p. 177.

\refis{3} K. Miyake, S. Schmitt-Rink, C.M. Varma, \prb
34, 6554, 1986.  D.J. Scalapino, E. Loh, J. Hirsch, \prb 34,
8190, 1986.

\refis{4} L. Genzel \etal, \prb 40, 2170, 1989. E. Seider,
\journal
Zeitschrift fur Physik B, 83, 1, 1991.

\refis{5} M. Gurvitch \etal, \prl 63, 1008, 1989.

\refis{6} K.J. Sun \etal, \prb 42, 2569, 1990.

\refis{7} M. Sera, S. Shamoto, M. Sato, \journal
Physica C, 185, 1335, 1991.

\refis{8} M-F. Xu \etal, \prb 37, 3675, 1988.

\refis{9} S.E. Stupp, W.C. Lee, J. Giapintzakis, D.M.
Ginsberg, \prb 45, 3093, 1992.

\refis{10} T. Shibauchi \etal, \journal Physica C, 185, 1851,
1991.  E.J. Ansaldo, \etal, \journal Physica C, 185, 1889,
1991.

\refis{11} W.W. Warren, Jr. \etal, \prl 59, 1860, 1987.

\refis{12} W. Reichardt \etal, \journal Physica C, 162,
464, 1989.  H. Rietschel, L. Pintschovius, W. Reichardt,
\journal Physica C, 162, 1705, 1989.

\refis{13} J.S. Tsai, Y. Kubo, J. Tabuchi, \prl 58, 1979,
1987.

\refis{14} M.S. Hybertsen, E.B. Stechel, M.
Schluter, D.R. Jennison, \prb 41, 11068, 1990.

\refis{15} R. Zeyher, G. Zwicknagl, \journal Solid
State Communications, 66, 617, 1988.

\refis{16} R. Zeyher, \prb 44, 9596, 1991.

\refis{17} J.P. Toennies, \journal Physica Scripta, T19,
39, 1987.

\vfill\eject\vskip 0.5truein
\beginparmode
\noindent{\bf Figure 1.}  This is the Fermi surface for a one-band
model with nearest-neighbor and next-nearest-neighbor
hopping, taken from Ref. 14.  The dispersion relation is
$$\epsilon(k,\theta) = -2t({\rm cos}(k{\rm cos}\theta)+{\rm cos}(k{
\rm sin}\theta))
-4t'({\rm cos}(k{\rm cos}\theta){\rm cos}(k{\rm sin}\theta))$$ with
$t=430{\rm meV}$ and $t'=-70{\rm meV}$.  The electron
filling factor is $0.85$.  $\vec q_n$ connects two nodes on
the Fermi surface if the gap is of this d-wave form.

\noindent{\bf Figure 2.}  This is a plot of the threshold line in the
(100) direction for the Fermi surface depicted in Figure 1.
The solid line is the line for the d-wave gap $\Delta(\theta)
=\Delta_o\left[{\rm
cos}(k_F(\theta){\rm cos}\theta)-{\rm cos}(
k_F(\theta){\rm sin}\theta)\right]$
and the dashed line is for the
isotropic $\Delta(\theta)=\Delta_o$ case.

\endmode\vfill\supereject
\end